\documentclass[preprint,
prl,
aps]{revtex4}
\usepackage[utf8x]{inputenc}
\usepackage{graphicx}
\usepackage{amsmath}
\usepackage{amsfonts}
\begin{document}
\title{Exactly solvable model for self-assembly of hard core - soft shell particles at interfaces}
\author{ Alina Ciach and Jakub P\c ekalski }
\address{
Institute of Physical Chemistry, Polish Academy of Sciences,
Kasprzaka 44/52, 01-224 Warsaw, Poland}
 \begin{abstract}
 A lattice model with soft repulsion followed by attraction is developed for a monolayer of 
 hybrid core-shell particles self-assembling
 at an interface. The model is solved exactly in one dimension. 
 One, two or three periodic structures and variety of shapes
 of the pressure-density isotherms may occur in different versions of the model. For strong interactions the isotherm
 consists of vertical segments separated by plateaus.
 The range of order depends  strongly on the strength of 
 attraction and on the density. Our results agree  with experimental observations. 
 \end{abstract}

 \maketitle

Hybrid hard-core soft-shell particles (HCSS) consisting of solid cores 
encapsulated in a cross-linked hydrogel network 
can self-assemble into ordered patterns on air-water or oil-water interfaces 
~\cite{vogel:12:0,nazli:13:0,volk:15:0,honold:15:0,geisel:15:0,rauh:16:0,karg:16:0}.
 Highly ordered arrays of particles with cores having desired properties
 can find applications in various fields, e.g. in surface patterning~\cite{rey:16:0}, photovoltaics~\cite{atwater:10:0}, 
 plasmonics~\cite{noginov:09:0}, sensing~\cite{willets:07:0} and
 emulsion stabilization~\cite{nazli:13:0}, and the question how to obtain desired 
 ordered patterns draws increasing attention.
 
 The patterns and the degree of order depend on the  core and  shell properties, 
 as well as on the surface pressure.  For pure
poly-N-isopropylacrylamide (PNIPAM) particles~\cite{nakahama:02:0} and for hybrid Au@PNIPAM
 particles with small Au cores~\cite{vogel:12:0},
similar patterns at the air-water interface and similar surface pressure - area isotherms were obtained. 
In both systems the particles form a hexagonal lattice. The surface pressure  $p$ increases with a moderate slope
for a large range of decreasing area; the moderate increase of  $p$ is followed by a  rapid increase, a 
 plateau and another rapid increase in a compressed monolayer. 
 In the case of silica@PNIPAM 
particles with relatively large silica cores 
adsorbed at the water-oil interface~\cite{rauh:16:0}, 
 more complex 
patterns are formed at large pressure. Moreover,
the  surface pressure - area isotherms 
are quite different than in the cases of the pure PNIPAM and Au@PNIPAM particles.
The isotherms of the silica@PNIPAM particles
have a characteristic shape of alternating segments with very large and quite 
small slope. The pressure range at the steep parts of $p(\eta)$ depends on the shell thickness.
Notably, the nearly vertical segments of  $p$ occur
for the area fraction of the particles, $\eta$, corresponding
to quite small area fraction of their cores. A natural question arises why for a few values of  
 $\eta$ a very large increase of $p$ is required  to induce any increase of the area fraction, while for
 area fractions intermediate between these distinguished values the compressibility of the monolayer is very large.  
 The fundamental question if the different patterns correspond to thermodynamically stable phases, and
  the plateaus  indicate phase transitions
 remains open.
 
 To the best of our knowledge there have been no attempts to develop a theory for the self-assembly of the 
 HCSS particles adsorbed at an interface that would  
 guide experimental studies. 
 Here
we construct a  coarse-grained model based on experimental observations.
 Model systems with two dimensional (2D) patterns can be studied either 
by simulations or by approximate theoretical methods. The simulations
of self-assembling systems are strongly influenced by finite size effects, and 
in theoretical studies the approximations may lead to incorrect results. 
In order to avoid possible inaccuracies resulting from approximations, we 
 introduce a one-dimensional (1D) lattice model that can be solved exactly.

To construct a coarse-grained  model for the HCSS particles adsorbed at an interface, 
we take into account that at  low area fraction the particles form a hexagonal lattice, 
 and  when $\eta$ further decreases, 
then the ordered structure remains unchanged, and
coexists with voids~\cite{rauh:16:0}. This suggests an attractive potential with 
a well-defined minimum at the separation $r=\sigma_a$. 
One source of the attraction may be 
a water ``cap''  formed above  the hydrophilic polymers grafted on the nanoparticle
 ~\cite{rauh:16:0}.
The caps lead to undulated interface with increased area, and this area increase is larger for particles
at large separations than for 
particles whose shells overlap. 
 The minimization of the surface-tension contribution to the free energy leads to 
effective attraction between the particles 
when their distance is larger than their diameter $\sigma_a$. Attraction might result from the van der Waals 
interactions between the monomers too~\cite{chremos:16:0}. 
On the other hand, when the shells of the two particles overlap,
they repel each other. The repulsion increases with decreasing distance between the particles.
Because the polymeric chains become compressed near the hard cores, the distance of the closest 
approach of two core-shell particles, $\sigma$, is larger than
the diameter of the solid core,  
and depends on the number and length of the grafted polymeric chains, and on cross-linking.
 
Based on the above facts, we conclude that the
effective interaction between the particles consists of the
steric repulsion at the distances $r$ smaller than  $\sigma$, next
of a soft repulsion for $\sigma<r<\sigma_a$, 
and finally of an attraction for $r>\sigma_a$~\cite{rauh:16:0}. 
\begin{figure}[h]
\centering
\includegraphics[scale=1]{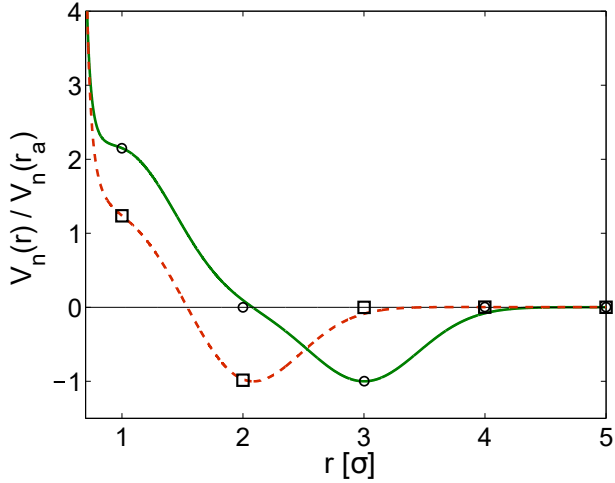}
\caption{Schematic illustration of the effective potential between the HCSS particles
adsorbed at an interface.
In the lattice model only discrete values of $r/\sigma$, indicated by the symbols, are considered; 
dashed and solid lines correspond to models I and II respectively. 
}
\label{fig.sketch}
\end{figure}

In Ref. ~\cite{rauh:16:0} monolayers of three types of HCSS particles with the same silica cores
and 
diameters $\sigma_a\sim 450nm,680 nm$ for the smallest and the largest shell
were investigated.
Based on the histograms for the nearest-neighbor distance in monolayers under large pressure~\cite{rauh:16:0:ESI}, we can
 expect that in each case $\sigma_a/\sigma \sim 2-3$, and the potential has a shape shown
 schematically in  Fig.\ref{fig.sketch}. 

We assume that the incompressible cores of the particles occupy lattice sites with the lattice constant $\sigma$.
The steric repulsion leads to forbidden multiple occupancy of the lattice sites.
We assume that the nearest-neighbors on the lattice repel 
each other with the strength $J_r>0$ (soft shell). In order to compare shells with different thicknesses,
we consider two variants of the model. In the first one the second neighbors attract each other
and the corresponding potential has the strength $-J_a$, with $J_a>0$.
For larger separations the effective potential vanishes. 
On the 1D lattice the  potential is given by 

\begin{equation}
\label{V1}
V_n(\Delta{ x}) = \left\{ \begin{array}{ll}
J_r & \textrm{for $|\Delta{ x}| = 1$},\\
-J_a & \textrm{for $|\Delta{ x}| = n$},\\
0 & \textrm{otherwise,}
\end{array} \right.
\end{equation}
with $n=2$ (model I). 
The positions and distances between the particles on the 1D lattice are 
denoted by $x=r/\sigma$ and  $\Delta x=\Delta r/\sigma$,
and take integer values. 
In the second variant of the model  the  potential changes sign for the second neighbors,
and is given by Eq.(\ref{V1}) with $n=3$ (model II).

We consider an open system with fixed chemical potential
of the particles, $\mu_p$, and fixed temperature $T$. 
We assume that the lattice consists of 
$L$ sites labeled from $1$ to $L$,
and consider periodic boundary conditions ($L+1\equiv 1, 0\equiv L$).
We  introduce an occupation operator $\hat\rho(x)$ which is equal to 1 or 0 when the site 
$x$ is occupied or not, respectively. 
Hence, the configuration of the system is given by
$\{\hat\rho(x)\}\equiv(\hat\rho(1),...,\hat\rho(L))$. Since each site can be either occupied or empty, 
there are $2^L$ configurations, and each of them occurs with 
the probability 
\begin{equation}
\label{probability}
P[\{\hat\rho(x)\}]=\frac{e^{-\beta H[\{\hat\rho(x)\}]}}{\Xi},
\end{equation}  
where
\begin{equation}
\label{Xi}
 \Xi=\sum_{\{\hat\rho(x)\}} e^{-\beta H[\{\hat\rho(x)\}]}
\end{equation}
 is the Grand Partition function, $\beta=(k_BT)^{-1}$, $k_B$ is the Boltzmann constant
and $H$ is the thermodynamic Hamiltonian which
contains the energy and the chemical potential term,
\begin{equation}
\label{H}
  H [\{\hat\rho\}] = \frac{1}{2} \sum_{x=1}^L\sum_{x'=1}^L \hat{\rho}( x) V ( x-x')\hat{\rho}(x')
- \mu \sum_{x=1}^L \hat{\rho}( x).
  \end{equation}
The energy of adsorption of a single particle at the interface, $h$, is included in $\mu=\mu_p+h$.  
The grand potential is given by
\begin{equation}
\Omega=-pL=-k_BT\ln \Xi
\label{Omega}
 \end{equation}
where $p$ is the 1D pressure.
We  also calculate 
 the dimensionless number density 
 $\rho=\langle \hat\rho(x)\rangle $ (length fraction of the cores) and the correlation function,
\begin{equation}
\label{correlation}
g(\Delta x)=\frac{\langle \hat\rho(x)\hat\rho(x+\Delta x)\rangle}
{\langle \hat\rho(x)\rangle\langle \hat\rho(x+\Delta x)\rangle},
\end{equation}
 with the probability distribution (\ref{probability}).
Because of translational invariance, $\langle \hat\rho(x)\rangle$ is independent of $x$, and $g$ depends
only on $\Delta x$. 

In the first step we  determine the ground state (GS), i.e. the structure at $T=0$. 
For $T=0$, the grand potential 
reduces to the minimum of $H [\{\hat\rho(x)\}]/L$. We  find  the minimum of $H [\{\hat\rho(x)\}]/L$
by comparison of $H [\{\hat\rho(x)\}]/L$ for empty and fully occupied lattice, and for 
different periodic structures.

 In the second step we consider $T>0$, using 
the transfer matrix method~\cite{pekalski:13:0}.
For the interaction range $n$
\begin{eqnarray}
\label{Xila}
 \Xi=Tr {\mathbb T}^{L/n} =\sum_{i=1}^{2^n}\lambda_i^{L/n},
\end{eqnarray}
where $\mathbb T$ is the  $2^n\times 2^n$
transfer matrix, and $\lambda_i$ are the eigenvalues of  $\mathbb T$ numbered such that 
$|\lambda_i|\ge|\lambda_{i+1}|$~\cite{pekalski:13:0}.
In the thermodynamic limit 
\begin{eqnarray}
\label{om}
 p=
 \frac{k_B T}{n}\ln \lambda_1.
\end{eqnarray}
 The matrix $\mathbb T$ is a finite matrix with positive elements,
 therefore $\lambda_1$ is non-degenerate. 
Thus,  for given $\mu$  Eq.(\ref{om}) yields a unique value of pressure, $p(\mu)$.
The average density $\rho(\mu)$ can be expressed in terms of 
the matrix $\mathbb P$ transforming  $\mathbb T$
to its eigenbasis~\cite{pekalski:13:0}. By eliminating $\mu$ from   $p(\mu)$ and $\rho(\mu)$, we obtain $p(\rho)$. 
The correlation function can be expressed in terms of
$\mathbb P$ and $\lambda_i$~\cite{pekalski:13:0}.
For large separations the correlations decay exponentially, with the 
 correlation length $\xi$ given by~\cite{pekalski:13:0}
\begin{equation}
\label{xi}
 \xi=n\Bigg[\ln \Big(\frac{\lambda_1}{|\lambda_2|}\Big)\Bigg]^{-1}.
\end{equation}
When $\lambda_2$ is real and positive, the decay is monotonic.
Because $\mathbb T$  is not symmetric,  
 pairs of complex-conjugate eigenvalues for $i>1$  may occur.  
The pair of complex-conjugate eigenvalues for $i=2,3$  leads to oscillatory
decay of correlations, with  the asymptotic form  for $x\gg 1$ 
\begin{eqnarray}
\label{corr1}
  g(x)= {\cal A}_i e^{-x/\xi}\cos\big(
x\lambda +\theta_i
\big) +1,
\end{eqnarray}
where the wave number $\lambda$ is the phase of the complex eigenvalue $\lambda_2=|\lambda_2|e^{i\lambda}$,
and   ${\cal A}_i$ and  $\theta_i$ 
depend on  $\mathbb P$
and on $i=mod(x, n)$~\cite{pekalski:13:0}.

 Let us start by discussing the GS. 
It turns out that in model I only one periodic structure with  alternating empty and occupied sites,
$\bullet$o$\bullet$o$\bullet$o..., and the unit cell ($\bullet$o)
may occur. By $\bullet$ we denote an occupied site, i.e. the uncompressible core of the particle.
The GS of model I is shown in Fig.2a
in variables $(\bar\mu=\mu/J_a,\bar J=J_r/J_a)$. 
\begin{figure}[h]
\centering
\includegraphics[scale=0.7]{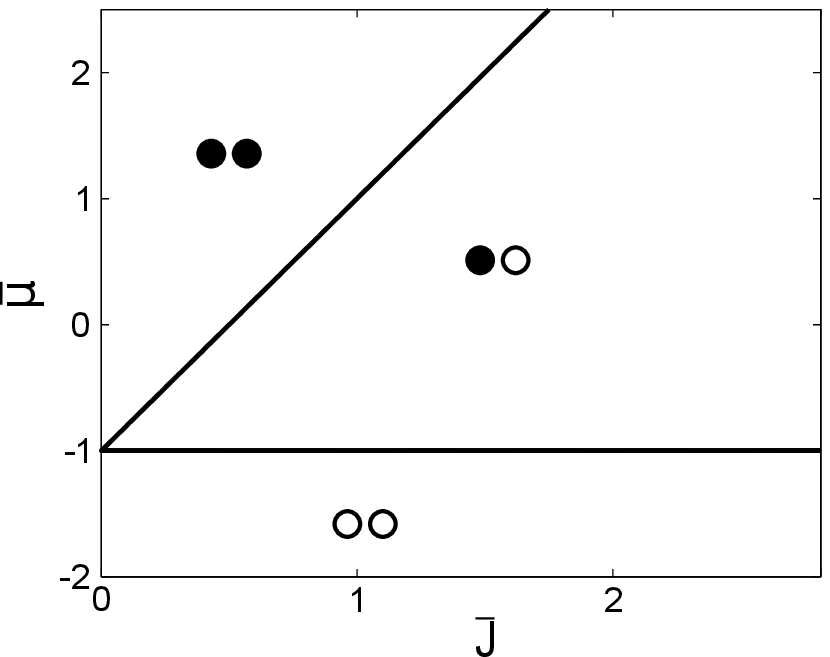}
\includegraphics[scale=0.7]{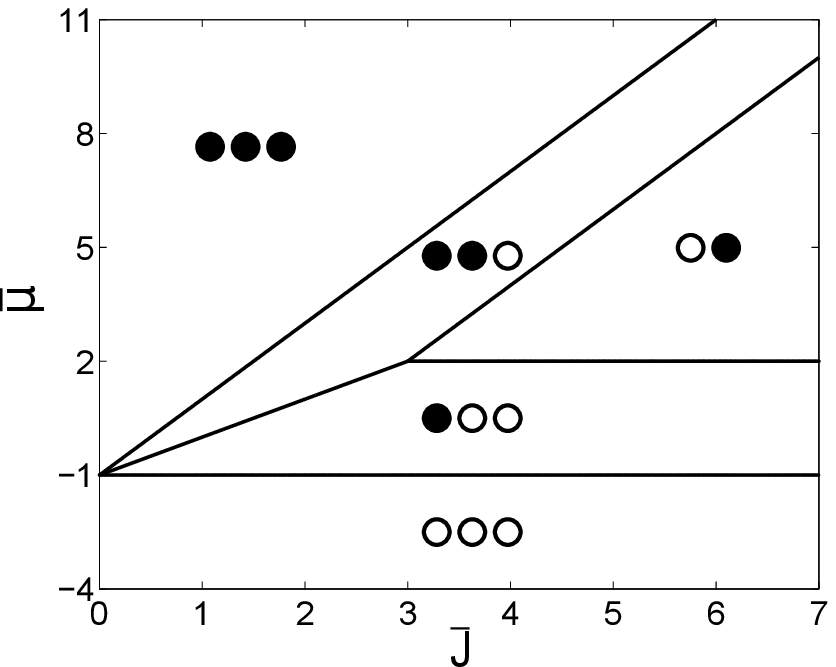}
\caption{ (a)  GS of  model I 
and (b) GS of model II. The coexistence lines are: (a) $\bar \mu=-1$ for
empty lattice - ($\bullet$o) and $\bar\mu=2\bar J -1$ for ($\bullet$o) - full occupancy;
 (b) $\bar\mu=-1$ for
empty  lattice  - ($\bullet$oo); $\bar\mu=\bar J-1$ for ($\bullet$oo) - ($\bullet\bullet$o); $\bar\mu=2$ for
($\bullet$oo) -($\bullet$o); $\bar\mu =2\bar J -4$  for ($\bullet$o) -  ($\bullet\bullet$o)
and finally  $\bar \mu= 2\bar J-1$ for ($\bullet\bullet$o) - full occupancy.
$\bar\mu=\mu/J_a$ and $\bar J=J_r/J_a$.
}
\label{fig.GS}
\end{figure}
In model II, three
periodic structures may occur  (Fig.\ref{fig.GS}b).
In the structure with  $\rho=1/3$, 
an occupied site is followed by two empty sites, 
and the unit cell is
($\bullet$oo). In the structure with $\rho=2/3$, two occupied
sites are followed by one empty site, 
and the unit cell is ($\bullet\bullet$o). 
The 
phase ($\bullet$o) with $\rho=1/2$
occurs only when $\bar J\ge 3$. 
To distinguish the  densities of the  periodically ordered GS structures, we use the notation $\rho_p$,
 i.e. $\rho_p=1/3,1/2,2/3$.

\begin{figure}[h]
\centering
\includegraphics[scale=1]{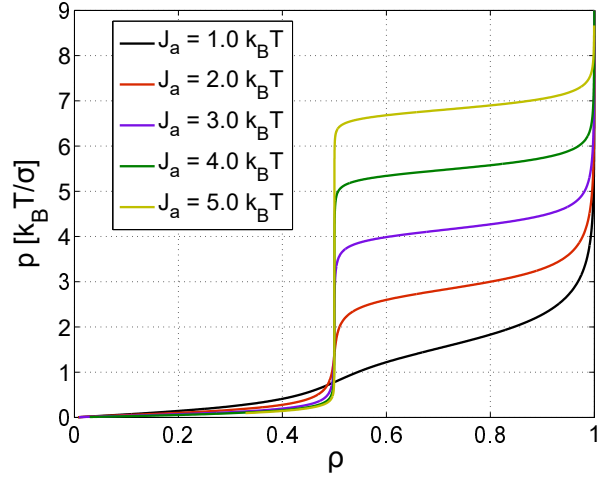}
\caption{The pressure-density isotherms in model I for  $\bar J=2$. 
 }
\label{fig.pzeta_1}
\end{figure}

\begin{figure}[h]
\centering
\includegraphics[scale=1.5]{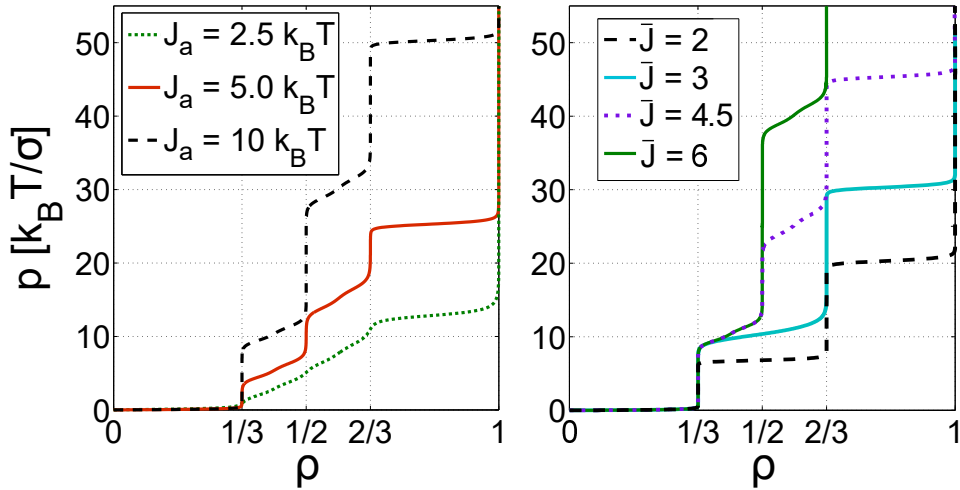}
\caption{The pressure-density isotherms in model II. (a) $\bar J=5$ (b) $J_a=10k_BT$.
 }
\label{fig.pzeta_2}
\end{figure}

 The results for $p(\rho)$ are shown in  Figs.\ref{fig.pzeta_1},\ref{fig.pzeta_2}.
 In both models,  nearly vertical segments for $\rho\approx\rho_p$ are separated by 
 nearly horizontal segments for $\rho\ne \rho_p$, when the  interactions are sufficiently strong.
 For  model I, there exists only one segment of the $p(\rho)$ curve with a very large slope (apart from $\rho\to 1$),
 consistent with the single periodic phase at $T=0$. 

In model II,  the nearly vertical segments of $p(\rho)$ are present 
for $\rho\approx1/3,2/3$, consistent with the GS structures 
($\bullet$oo) and ($\bullet \bullet$o).
When  $\bar J> 3$, a 
third ``step'' at $\rho=1/2$ appears (Fig.\ref{fig.pzeta_2}b). For 
  fixed $\bar J$, the  pressure range for $\rho\approx\rho_p$ increases
  with increasing $J_a$ (Fig.\ref{fig.pzeta_2}a), whereas for fixed $ J_a/(k_BT)$ and increasing $\bar J$ 
the  pressure range increases significantly  
only at the central step. This behavior is consistent with the GS, where  the  ($\bullet$o) phase
is stable for the range of
the chemical potential that increases for increasing $\bar J$.
Quite surprisingly,  when
 $J_a/(k_BT)$  is fixed and $\bar J>3$, 
  $p(\rho)$   is nearly independent of $\bar J$ for $\rho<1/2$. 

\begin{figure}[h]
\centering
\includegraphics[scale=1.5]{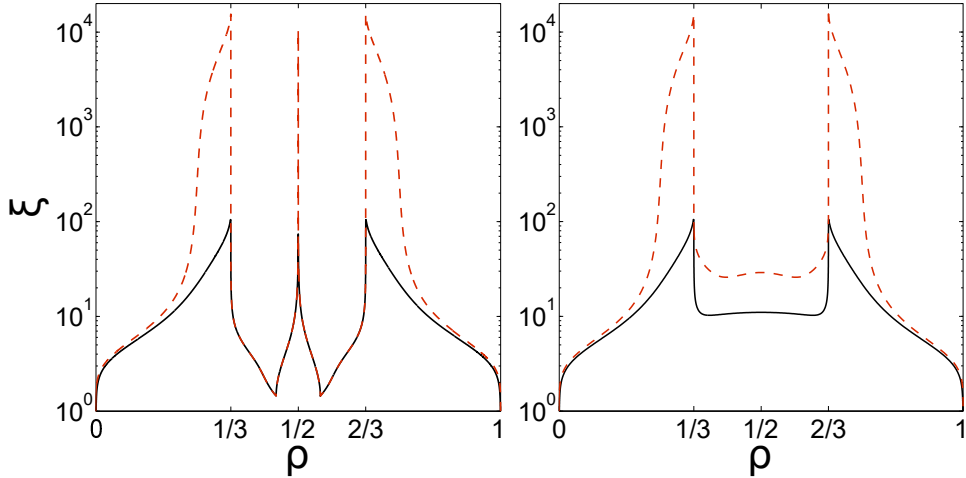}
\caption{The correlation length $\xi$ (in  $\sigma$ units)
in model II for $\bar J=5$  (left panel) and for $\bar J = 2$ (right panel)
with $J_a=5k_BT$ (black solid line) and $J_a=10k_BT$ (red dashed line).
The number density $\rho$ is dimensionless.
}
\label{fig.cor1}
\end{figure}

We have found that  $g(x)$ is given by (\ref{corr1}), with the period of oscillations $2\pi/\lambda\approx 2$ for 
$\rho\approx 1/2$ and $2\pi/\lambda\approx 3$ otherwise, in agreement with the GS structures. 
The correlation length is very large for $\rho=\rho_p$, and increases rapidly 
for increasing $J_a$ (Fig.\ref{fig.cor1}). However, 
 when $\rho$ departs slightly from $\rho_p$, $\xi$ decreases by orders of magnitude
 and becomes independent of $J_a$ for $1/3< \rho< 2/3$. 
 Slight deviations from $\rho_p$ lead to dramatic decrease of order in this range of density.
 For $\rho<1/3$ or $\rho>2/3$,
$\xi$ decreases much more slowly for $\rho$ departing form $\rho_p$. Moreover, $\xi$ 
depends very strongly on $J_a$ and very weakly on $\bar J$ for $\rho<1/3$ or $\rho>2/3$.

We have obtained remarkably rich behavior from the very simple model (\ref{V1}). One, two or three periodic
structures with the corresponding vertical segments of $p(\rho)$ can occur, depending on the ranges and strengths
of the repulsive and attractive parts of the potential. 
Phase transitions
and   long-range order are absent for $T>0$ in  1D models with finite range of interactions. 
Our results obtained in such a model show that
the plateaus in the 
$p(\rho)$ curve can be present even in the absence 
of true thermodynamic phase transitions. The short-range order in the disordered phase mimics 
the long-range order of the phase
stable at lower $T$ (at $T=0$ in 1D), and the range of this order can be orders of magnitude larger than $\sigma$. 

The strength of attraction plays a key role in  formation of ordered patterns and in the shape of the pressure-area 
isotherm for small densities. 
In contrast,
for densities larger than the close packing density of the soft particles, the repulsion
determines the shape of $p(\rho)$ and the range of order.

Our results agree surprisingly well with experiments. 
 The isotherm obtained in Ref.\cite{vogel:12:0} for small Au cores  has the shape that agrees with
 the isotherm shown in Fig.\ref{fig.pzeta_2}a for $J_a=2.5k_BT$. For increasing density, the sequence of 
   very small,  moderate, very large,  very small, and again very large slope of the pressure is found in both cases.
 The alternating steep and shallow segments obtained in Ref.\cite{rauh:16:0} for larger silica cores 
 agree with the curves obtained in our model for stronger attraction. 
 In Ref.~\cite{rauh:16:0} and in our model (Fig.\ref{fig.cor1}),  the  samples with density
 larger than the density of close packing are less ordered.
Increase of the particle diameter $\sigma_a$ leads 
 to increase of both, the range of order for 
 small area fractions, and the pressure range at the corresponding vertical segment of the isotherm~\cite{rauh:16:0}.
 In our model such behavior is found for increasing $J_a$ (Figs.\ref{fig.cor1},\ref{fig.pzeta_2}a). 
 This observation indicates that the attraction increases with increasing $\sigma_a$, and
 supports the conjecture that the attraction in Ref. \cite{rauh:16:0} results from 
  the surface-tension contribution to the free energy. 
 
   Our results indicate that if ordered structures are desired, one should try to increase the strength 
 of the attractive part of the interactions, and
  choose area fraction of particles approaching the close-packing density from below. 
  For denser systems the density should be fixed with extremely high precision to achieve large correlation length.
 
 Models with repulsive shoulder followed by attractive well 
 were studied before in different contexts~\cite{hemmer:70:0,kincaid:76:0,oliveira:06:0,lomba:07:0}.
  In particular, multiple phase transitions\cite{hemmer:70:0,kincaid:76:0} and water
  anomalies were obtained \cite{oliveira:06:0,lomba:07:0}. 
  Our results show that  a potential of this kind
  (Fig.\ref{fig.sketch})
  is also able to reproduce the main features of the HCSS particles self-assembling at interfaces. 
  
The isotherms  very similar to Fig.\ref{fig.pzeta_1} were obtained for
  the 1D  model  with short-range attraction and long-range repulsion (SALR)~\cite{pekalski:13:0}, 
  and for the  1D model of aqueous solution of 
  amphiphilic molecules~\cite{pekalski:14:1}.
  In  model I and in Ref.~\cite{pekalski:13:0,pekalski:14:1} a single phase with periodic arrangement 
  of the particles, clusters or micelles was found in the GS. The periodic order 
 is reflected in a very large slope of the 
pressure for the density or concentration optimal for the periodic 
structure, 
independently of the kind of ordering objects and the source of competing interactions.   
 Such universal properties can be correctly predicted by generic models,  and models like the one introduced in this work can
guide future experiments.

We acknowledge the financial support by the National Science Center grant  2015/19/B/ST3/03122. 
JP acknowledges the financial support by the National Science Center under Contract Decision No. DEC-2013/09/N/ST3/02551.

 \end{document}